\documentclass{article}
\usepackage{spconf,amsmath,graphicx,bm,cite,setspace}
\usepackage{lipsum}
\usepackage{multirow}

\usepackage[]{algorithm2e}

\title{On Evaluating CNN representations for Low resource medical image classification }
\name{Taruna Agrawal, Rahul Gupta, Shrikanth Narayanan}
\address{Signal Analysis and Interpretation Lab, University of Southern California}

\begin{document}
\ninept
\maketitle
\begin{spacing}{1.0}
\begin{abstract}
Convolutional Neural Networks (CNNs) have revolutionized performances in several machine learning tasks such as image classification, object tracking, and keyword spotting.
However, given that they contain a large number of parameters, their direct applicability into low resource tasks is not straightforward.
In this work, we experiment with an application of CNN models to gastrointestinal landmark classification with only a few thousands of training samples through transfer learning.
As in a standard transfer learning approach, we train CNNs on a large external corpus, followed by representation extraction for the medical images.
Finally, a classifier is trained on these CNN representations.
However, given that several variants of CNNs exist, the choice of CNN is not obvious.
To address this, we develop a novel metric that can be used to predict test performances, given CNN representations on the training set. 
Not only we demonstrate the superiority of the CNN based transfer learning approach against an assembly of knowledge driven features, but the proposed metric also carries an 87\% correlation with the test set performances as obtained using various CNN representations.
\end{abstract}

\renewcommand{\thefootnote}{\fnsymbol{footnote}}

\vspace{-2mm}

\section{Introduction}
\label{sec:intro}
\vspace{-2mm}

Recent advances in the design of Convolutional Neural Networks (CNNs) has led to state of the art performances in several tasks, including image classification \cite{krizhevsky2012imagenet}, object detection \cite{ren2015faster} and object tracking \cite{kristan2015visual}.
CNNs can be viewed as models that extract features from raw images using convolution and pooling operations, followed by a classification using fully connected layers \cite{krizhevsky2012imagenet}.
However, training these models typically requires large amount of samples, given the large number of trainable parameters.
Transfer learning is a promising approach in such cases, wherein CNN models are pre-trained on larger unrelated corpora, followed by a fine-tuning on the task of interest. 
The task of interest in this paper is the classification of gastro-intestinal (GI) tract images, given a few hundred training samples per class.
A vanilla classification approach in this case would be extraction of a selected set of features, followed by learning a classifier.
First, we establish the superiority of transfer learning approach using CNN models learnt on larger unrelated corpora against the vanilla classification approach.
However, given that several variants of CNN architectures exist, it is not evident which CNN representation will yield the best performance.
To address this, we also propose a metric that can inform the choice of CNN architecture.

{\bf Previous work:} CNNs have revolutionized research in several fields such as image classification/detection \cite{krizhevsky2012imagenet}, automatic speech recognition \cite{abdel2013exploring} and natural language understanding \cite{kim2014convolutional}.
CNNs typically perform a set of convolution and pooling operations, variants of which have been proposed by several researchers \cite{howard2017mobilenets,he2016deep}.
These variants are typically developed taking into consideration the task at hand.
A few examples with custom CNN design include acoustic modeling for low resource languages \cite{alumae2016improved}, object and action classification \cite{oquab2014learning} and remote sensing \cite{xie2015transfer}.
On the other hand, medical image classification \cite{antonie2001application} requires assignment of medical images (drawn from real world patients) to a medical landmark, phenomenon or a disease and often, obtaining large amounts of training data can be challenging.
A few approaches for medical image classification include the use of decision trees \cite{rajendran2010hybrid}, k-nearest-neighbors \cite{ramteke2012automatic} and support vector machines \cite{fan2005classification}.
Researchers have also applied CNNs for medical image classification using training from scratch \cite{li2014medical} as well as transfer learning \cite{shin2016deep}. 
In their work, Tajbakhsh et al. \cite{tajbakhsh2016convolutional} address questions regarding the choice between full training versus fine tuning based on empirical performance evaluation.
Shin et al. \cite{shin2016deep} also simplify existing CNN architectures to reduce the number of parameters for training on medical imaging dataset.
Recently the MediaEval challenge \cite{michael2017multimedia} garnered further interest in medical image classification, with proposals to use CNN based classifiers \cite{agrawalscl, petscharnig2017inception}.
All the above papers report performances using a CNN, however they fall short of describing a process that can inform selection of a CNN variant appropriate for the task at hand.
First, we obtain performances on the KVASIR dataset \cite{Pogorelov:2017:KMI:3083187.3083212} using transfer learning based approach. 
Using these results as an empirical testbed, we propose a metric that can predict performances on the test set.
The goal of this metric is to inform the decisions regarding the choice of CNN architecture for transfer learning.

For the task of GI landmark classification, we first establish the performances using two kinds of approaches (i) a kitchen sink feature extraction and classifier training and, (ii) extracting mid-level CNN representations followed by classification layer fine-tuning.
We observe that the CNN based transfer learning based approach obtains significantly better test performances on the dataset of interest (described in Section~\ref{sec:dataset}) for a majority of CNN variants.
However, in real world, choosing a CNN representation based on test performances is not feasible.
To address this issue, we propose a metric that can be computed using the training set to predict performance on the test set.
We aim for a one shot metric estimation that is robust to the absence of large training sets.
We propose a metric whose computation entails projecting the training data-points into a lower dimension, followed by estimation of class confusions in the projected space. 
Given the various feature representations, the trends predicted by the proposed metric carries a correlation coefficient of 0.87 with the actual test accuracies.




\vspace{-2mm}

\section{Dataset}
\label{sec:dataset}
\vspace{-2mm}

We use the KVASIR dataset \cite{Pogorelov:2017:KMI:3083187.3083212} in our experiments.
The dataset consists of 8000 images, equally drawn from eight different GI anatomical landmarks: (i) esophagitis, (ii) normal z-line, (iii) ulcerative-colitis, (iv) normal-pylorus, (v) polyps, (vi) dyed-lifted-polyps, (vii) dyed-resection-margins and, (viii) normal-cesum.
The size of these images ranges between 720x576 to 1920x1072, each annotated by a professional endoscopist. 
In order to perform experiments, we use a training and testing set partition suggested in \cite{michael2017multimedia}, with 4000 instances in each partition.
The objective behind this dataset collection is to aid early discovery of lesions, that can prevent cancer progression. 
More information regarding the dataset can be obtained from \cite{Pogorelov:2017:KMI:3083187.3083212, michael2017multimedia}.

\vspace{-2mm}
\section{Classification Methodology}
\label{sec:method}
\vspace{-2mm}

We obtain a representation for each GI image in the KVASIR dataset using two strategies:
(i) a baseline kitchen sink feature extraction strategy and, (ii) feature representations obtained using CNNs trained on external corpora. 
These representations are then used to train a classifier on the available training data.
We describe the feature extraction below, followed by the classification setup.

\begin{table}[t]
\caption{Brief description of features used as a baseline.}
\begin{center}
\begin{tabular}{ @{}lr@{} } 
\hline
{\bf Feature} : { Description} \\ \hline
{\bf Joint Composite Descriptor}: Carries color and texture \\
information in a compressed format\\ \hline
{\bf Tamura}: Features corresponding to human visual perception: \\  coarseness, directionality, line-likeness, regularity, and roughness\\ \hline
{\bf ColorLayour}: Spatial distribution of color in the image\\ \hline
{\bf EdgeHistogram}: Capture edge distribution in the image \\ \hline
{\bf AutoColorCorrelogram}: Capture color correlation information\\
in the image\\ \hline
{\bf Pyramid Histogram of Oriented Gradients}: Quantifies spatial \\ layout and local shapes within the image \\ \hline
\end{tabular}
\label{tab:features}
\vspace{-8mm}
\end{center}
\end{table}

\vspace{-2mm}
\subsection{Baseline: kitchen sink feature extraction strategy}
\vspace{-1mm}

In this strategy, we use an assembly of knowledge driven features (as opposed to the data driven feature representations extracted in CNNs).
We use a set of baseline features, as shown in Table~\ref{tab:features}.
These features were motivated by Pogorelov et al. \cite{Pogorelov:2017:KMI:3083187.3083212,michael2017multimedia} for application to the KVASIR dataset.
These features are global descriptors of the images and are designed in a knowledge driven manner to capture a specific property of the images.
The dimensionality of baseline feature representation is 1179.

\vspace{-2mm}
\subsection{CNN based feature extraction}
\vspace{-1mm}

In this strategy, we initially train CNN models on an external unrelated dataset, the ImageNet dataset \cite{krizhevsky2012imagenet}.
We then obtain image representations yielded by these networks in the penultimate layer (layer right before the output layer) for the KVASIR dataset.
We scale each image in the KVASIR dataset to a size of 224 $\times$ 224, equal to the size of images in the ImageNet dataset.
These scaled images are then fed to the CNNs and we apply global average pooling to the outputs of last convolutional layer in each of the CNNs. 
We test five CNN architectures in our experiments, as described in Table~\ref{tab:cnns}.

\begin{table*}
    \centering
            \caption{Features representations used in our experiments. We also present the accuracies for the experiment presented in Section~\ref{sec:exp}. Results for VGGNet, Inception-V3, XceptionNet and MobileNet are significantly better than the baseline (binomial proportions test, p-value $<1\%$) }
    \begin{tabular}{@{}lll@{}l@{}}
    \hline
        \bf Features & \bf Description & \bf Feature  & \bf Accuracy\\ 
         &  &  \bf dimensionality & \\ \hline
        Baseline & See Table~\ref{tab:features} & 1179 & 71.6 \\ \hline 
        VGGNet \cite{simonyan2014very} & 16 layer architecture, uses $3\times3$ convolution $2\times2$ pooling throughout the network. & 512 & 80.1 \\
         ResNet50 \cite{he2016deep} & 50 layer networks with shortcut connections. & 2048 &61.1 \\
         Inception-V3 \cite{szegedy2016rethinking} & Performs convolution with filters of dimensionality $1\times1$, $2\times2$ and $3\times3$ & 2048 &75.6\\
         XceptionNet \cite{chollet2016xception} & Extension of the Inception architecture with standard inception modules  & 2048 & 80.8\\
          &  replaced by depthwise separable convolutions &  \\
         MobileNet \cite{howard2017mobilenets} & Uses \textit{depthwise separable convolution} to build light weight deep neural networks & 1024 & 81.7 \\ \hline
        Combined & & & 83.8 \\ \hline

    \end{tabular}
\vspace{-5mm}
\label{tab:cnns}
\end{table*}

\vspace{-2mm}
\subsection{Classification setup}
\label{sec:exp}
\vspace{-1mm}

After obtaining feature representations for an image, we train a multi-class Support Vector Machine (SVM) classifier for classifying image to one of the eight GI landmark labels.
In case of feature representations obtained from CNNs, this training can also be seen as pre-training the CNNs on large datasets and fine tuning the final classification layer using a hinge loss on the KVASIR dataset.
The hyper-parameters for the SVM classifier (kernel and box-constraint) are tuned using a five fold inner cross validation on the training set.
We present the classification results in the next section.

\begin{figure}[t]
\centering
\includegraphics[scale=0.4]{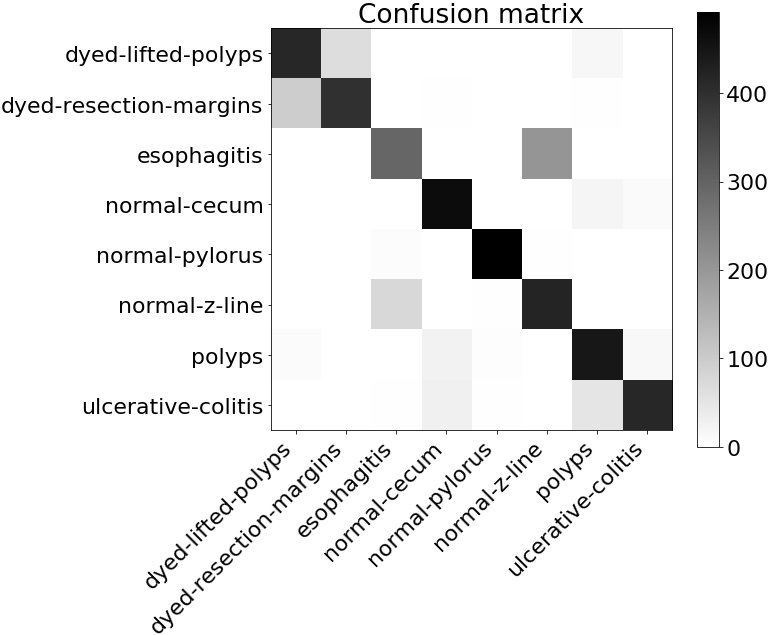}
\vspace{-3mm}
\caption{Confusion matrix for a system trained on representation combined from each source. Bar on the right indicates absolute count in each class.}
\vspace{-3mm}

\label{fig:confusion_matrix}
\end{figure}

\vspace{-2mm}
\subsection{Classification Results}
\label{sec:results}
\vspace{-1mm}

Given that the classes are balanced in the training and testing partitions, we use accuracy as our evaluation metric.
Table~\ref{tab:cnns} presents the results for each image representation.
We also present results for a case where we concatenate features representations from all the sources.
In almost all the cases, representations yielded by CNN outperform the baseline features (except ResNet).
This indicates that data driven representations obtained on external corpora can outperform knowledge based features.
Research has shown that CNN tends to learn filters sensitive to geometrical patterns observed in the training data \cite{mahendran2015understanding}. 
Since the CNNs are initially trained on a large set of images, our results suggest that they learn to encode geometrical patterns, which yield better classification results over knowledge based features.
We also observe that the combined model performs the best, indicating that the features from various sources are complementary.
We also plot the confusion matrix for the system using combination of all features in Figure~\ref{fig:confusion_matrix}, as obtained on the testing partition.
The confusion matrix is indicative of the classes that the classifier tends to confuse (e.g. we observe a confusion between the dyed-lifted-polyps and dyed-resection-margins class).
We refer back to the confusion matrix for further analysis in Section~\ref{sec:tsne}.

Although a majority of CNN representations yield better results than the baseline features, the choice amongst the CNN representations is not obvious during system design and training.
Therefore, we need a mechanism to assess the discriminative capability of each representation during training.
We address this by proposing a metric to estimate the accuracy yielded by a given feature representation in the next section.

\vspace{-2mm}
\section{Estimating accuracy yielded by feature representations}
\vspace{-2mm}

Given a feature representation (baseline or extracted from a CNN), we propose a metric to estimate the accuracy yielded by a classifier trained on those features.
We design the metric such that it could be computed based on the training set.
Furthermore, we should be able to obtain it using a one shot computation as opposed to estimation methods such as inner cross-validation on the training set.
Such a method requires pre-selection of a classifier, hyper-parameter tuning and is computationally expensive. 
In particularly, on a small dataset as ours, results could vary from one cross-validation split to the other, leading to a noisy estimate.
We outline the computation for the proposed metric in Algorithm~\ref{algo1}.

\begin{algorithm}[t]
 \label{algo1}
 \caption{Algorithm to estimate the metric $A$ on training data. $A$ is expected to inform the choice of a feature representation.}

 {\bf Given}: Training data $d_n, n = 1,..,N$ with associated feature representations $\bm x_n \in R^D$ and label $y_n \in \{ 1,..,K \}$ ($D$ is feature dimensionality and $K$ is the number of classes)\;
 
 {\bf Step 1: Obtain transformation} $\bm z_i = f(\bm x_i)$, where $f$ is an embedding function for $x_i$\;
  
  {\bf Step 2: Modeling class probabilities}\;

 \For{$k = 1,..,K$}  
 {
 {{\bf Estimate PDF} $P_{k} (\bm z | y = k) = \mathcal{N} (\bm \mu_k, \bm \Sigma_k)$\;} {Where $\bm \mu_k = $ Mean($\bm z_n$) $\forall d_n$ with class $k$\;}
 {$\bm \Sigma_k = $ Co-variance($\bm z_n$) $\forall d_n$ with class $k$ \;}
 }

 {\bf Step 3: Estimating accuracies based on the PDF $P_{k} (\bm z | y = k)$}\;
 \For{$k = 1,..,K$}  
  {
  {Estimate accuracy for class $k$\;}
  $A_k = \int_{\substack {\bm z: {P_k(\bm z) > P_{k^\prime(\bm z)} \forall k^\prime \neq k}} } P_k(\bm z) \partial \bm z $
  }
  {\bf Step 4: Final accuracy estimate:} 
  $A =$ mean($A_1,..,A_K)$
\end{algorithm}

\begin{figure}[t]
\centering
\includegraphics[trim={1cm 0 2 1cm},clip,scale=0.5]{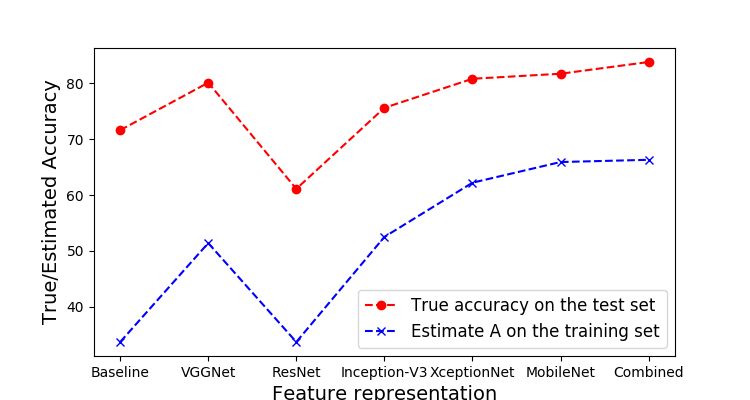}
\vspace{-5mm}
\caption{Plot comparing the accuracies obtained on the test set against the estimates $A$ obtained on each feature representation.}
\vspace{-2mm}

\label{fig:estimate}
\end{figure}

\begin{figure*}[t]
\centering
\includegraphics[scale=0.5]{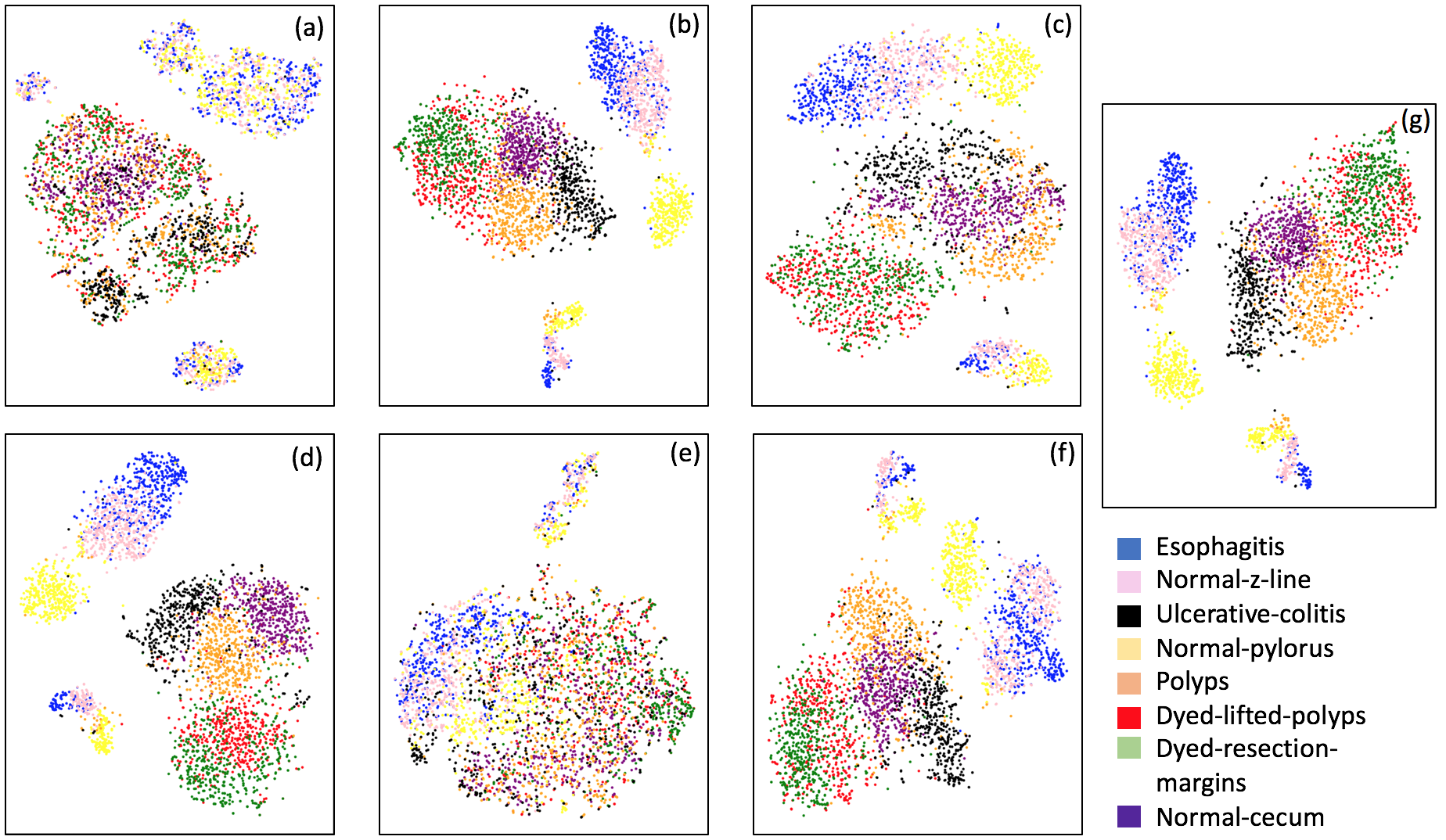}
\vspace{-4mm}
\caption{t-SNE plots obtained using various image feature representations: (a) Baseline features, (b) Inception-V3, (c) VGG-16, (d) MobileNet, (e) ResNet, (f) XceptionNet, (g) All representations combined. We chose a 2-dimensional projection for the ease of visual inspection.}
\vspace{-3mm}

\label{fig:single_results}
\end{figure*}

Our proposed algorithm first projects the feature representations from a high dimensional space into a lower dimension space using the projection function $f$.
This is followed by obtaining a Probability Distribution Function (PDF) $P_k(\bm z | y = k)$ of the projected data-points ($\bm z_n = f(\bm x_n)$) based on the class $k$ of their membership.
Projecting the data-points on to the lower dimensional space is desirable, as with limited data, the parameters estimation for class specific PDF, $P_k(\bm z | y = k)$, is more robust. 
We chose $P_k(\bm z | y = k)$ to be a Gaussian distribution.
In step 3, based on the estimated class distributions $P_k(\bm z | y = k)$, we compute the probability that a point $\bm z$ sampled from $P_k(\bm z | y = k)$ will yield highest PDF value from the same PDF.
We term this estimate as $A_k$ and it is integral of $P_k(\bm z | y = k)$ in the space spanned by $\bm z$ where $P_k(\bm z | y = k) >P_k^\prime(\bm z | y = k^\prime), \forall k \neq k^\prime $.
We average the $A_k$ from each class to obtain the final estimate $A$ (we chose averaging since the class distribution is uniform in the training set).
We expect the metric $A$ to be indicative of the accuracy obtained when using the feature representation $\bm x$.

We considered multiple lower dimension projection techniques such as Principal Component Analysis, auto-encoders and t-SNE.
Empirically, we observed that the t-SNE projections (in a 2-D space) yield good estimates for $P_k(\bm z | y = k)$ with different $\bm \mu_k, \bm \Sigma_k$ values for each class $k$.
$\bm \mu_k, \bm \Sigma_k$ estimates using other methods tend to be close, implying a high degree of overlap between class specific distributions $P_k(\bm z | y = k)$ in the projected space.
Next, we present our findings on the success of the proposed metric $A$ in predicting the test accuracy.

\vspace{-2mm}
\subsection{Results}
\vspace{-1mm}

Figure~\ref{fig:estimate} plots the accuracies obtained on the test set (as also presented in Table~\ref{tab:cnns}) against the estimate $A$ for each feature representation.
We note that although that the estimate is off by certain points, the metric $A$ captures the accuracy trend on the testing set.
We obtain a correlation of 87\% between $A$ and accuracies on the test set.
We argue that despite an error in prediction, high correlation with test accuracy is useful as it can inform what feature representation is likely to yield the highest accuracy.
We acknowledge that the absolute value of $A$ itself may be off the actual accuracy estimate. 
Another point to note is that the algorithm to compute the metric $A$ was not informed of the type of classifier (SVM) used in our experiments.
Therefore the estimation is performed independent of the final classifier and the associated hyper-parameters.



\vspace{-2mm}
\subsection{Analysis on t-SNE projections}
\label{sec:tsne}
\vspace{-1mm}

To further analyze the high correlation between the metric $A$ and test performance, Figure~\ref{fig:single_results} presents the t-SNE projections for each feature representation on the training set.
The class-wise distribution trends in Figure~\ref{fig:single_results} closely associate with the classification performances on the testing set.
We observe that in the case of baseline feature representations, different classes gets clustered together in the t-SNE projection plot.
The plot suggests that the t-SNE method deems images from different classes to be similar to each other, based on the baseline features.
This is coherent with the poor performance observed using the baseline features.
On the other hand, class separation is evident in the case of Inception-V3, VGG-16, MobileNet and XceptionNet CNN architectures.
The t-SNE representations using ResNet do not cluster as per the eight GI landmark classes, which is in line with the low performance observed using these representations.
Overall, the visual trends observed in Figure~\ref{fig:single_results} correspond to the actual test performances, explaining the success of metric $A$ in predicting test accuracies.


Another question we aim to answer is if we can predict the class confusion amongst the eight classes using the t-SNE analysis.
t-SNE plots in Figure~\ref{fig:single_results} present promising trends with this regards as well.
For instance in the confusion matrix (Figure~\ref{fig:confusion_matrix}), we observe that the class normal-pylorus has the least confusion with other classes.
In the t-SNE plot in Figure~\ref{fig:single_results}(g), we observe that this class occurs as a separate cluster by itself.
The clusters for three classes: ulcerative-colitis, normal-cecum and polyps, are close to each other, which does reflect as small amount of confusion amongst these three classes.
A large confusion is observed between dyed-lifted-polyps and dyed-resection-margins and, esophagitis and normal-z-line classes.
The clusters corresponding to these classes have fair amount of overlap in the t-SNE plots.
We note that one could obtain a pairwise class confusion metric between two classes $k$ and $k^\prime$ as $A_{k,k^\prime} = \int_{\substack {\bm z: {P_{k^\prime}(\bm z) > P_k{^*}(\bm z); k^* \neq k^\prime}} } P_k(\bm z) \partial \bm z $ (we integrate $P_k(\bm z)$ over the region where $P_{k^\prime}(\bm z)$ dominates).
We observed that this metric obtains mediocre performances in predicting class confusions (a correlation between 20\% - 50\%, depending upon the feature representation).
Since the development of this particular metric needs further research, we do not present the detailed results in this paper and consider this as an avenue for future research. 


\vspace{-2mm}
\section{Conclusion}
\label{sec:conclusions}
\vspace{-2mm}

Several variants of CNNs have been proposed in the past to address problems related to computer vision, speech recognition and natural language understanding.
We test their application on a medical imaging problem involving identification of GI landmarks given an image.
We use a set of baseline feature representations crafted to capture specific aspects of images as well as feature representations yielded by a set of five different CNN architectures.
Classifier trained on four out of five CNN representations outperform the baseline features.
Furthermore, we develop a novel metric to inform the choice of CNN architecture for obtaining these representations.
We observe that we can foretell the relative performance on the test set by using the proposed metric obtained on the training set.
We analyze that the success of the proposed metric stems from a robust lower dimension projection yielded by the t-SNE projections. 

In the future, we aim to perform further investigations on the transfer learning approach with CNNs.
As of now, we use representations as obtained from the penultimate layer.
However, intermediate representations may contain further complementary information.
One may also investigate additional low dimensional projection techniques to estimate performance on the testing set.
Future work may include decision on classifier design itself based on the t-SNE plots. 
For instance, a mixture of experts \cite{gupta2015mixture} model can be used to distinguish classes using a specific set of features, which otherwise carry a significant overlap in other sets of feature representations.

\end{spacing}

\newpage

\begin{spacing}{1.0}
\bibliographystyle{IEEEtran}
\bibliography{twitter}
\end{spacing}
\end{document}